\documentclass{article}
\usepackage{ijcai18}
\usepackage{times}
\usepackage{xcolor}
\usepackage{soul}
\usepackage[utf8]{inputenc}
\usepackage[small]{caption}
\usepackage{versions}
\excludeversion{old}
\excludeversion{fullversion}\includeversion{shortversion}
\excludeversion{sellers}

\usepackage{graphicx}
\graphicspath{{./graphics/}}

\usepackage{tabu}
\newcommand{\citet}[1]{\citeauthor{#1}~\shortcite{#1}}
\newcommand{\citep}{\cite}

\newcommand{\abs}[1]{\left|#1\right|}
\newcommand\range[2]{\in\{#1,\dots,#2\}}
\newcommand{\e}[1]{\exp{\left(#1\right)}}

\newcommand{\kx}{\ensuremath{k_{x}}}
\newcommand{\ky}{\ensuremath{k_{y}}}
\newcommand{\kmin}{\ensuremath{k_{\min}}}
\newcommand{\kmax}{\ensuremath{k_{\max}}}

\title{
 Double Auctions in Markets for Multiple Kinds of Goods 
}

\author{
Erel Segal-Halevi$^1$, 
Avinatan Hassidim,$^2$ 
Yonatan Aumann$^2$ 
\\ 
$^1$ Ariel University, Ariel 40700, Israel \\
$^2$ Bar-Ilan University, Ramat-Gan 52900, Israel
\\
erelsgl@gmail.com, avinatanh@gmail.com, yaumann@gmail.com
}

\newcommand{\ignore}[1]{}

\newcommand{\price}{\ensuremath{\mathbf{p}}}  % Optimal price
\newcommand{\priceo}{\ensuremath{\mathbf{p^O}}}  % Optimal price
\newcommand{\pricea}{\ensuremath{\mathbf{p^R}}}  % Right price
\newcommand{\priceb}{\ensuremath{\mathbf{p^L}}}  % Left price
\newcommand{\pricedelta}{\ensuremath{\mathbf{\Delta}}}  % Left price

\newcommand{\Byn}{\ensuremath{B_{y-}}}
\newcommand{\Bny}{\ensuremath{B_{+y}}}

\newcommand{\Bxn}[1]{\ensuremath{B_{x-}^{#1}}}
\newcommand{\Bnx}[1]{\ensuremath{B_{+x}^{#1}}}

\newcommand{\Bxs}[1]{\ensuremath{B_{x*}^{#1}}}
\newcommand{\Bys}[1]{\ensuremath{B_{y*}^{#1}}}

\newcommand{\Dxn}[1]{\ensuremath{D_{x-}^{#1}}}
\newcommand{\Dnx}[1]{\ensuremath{D_{+x}^{#1}}}

\newcommand{\Dny}[1]{\ensuremath{D_{+y}^{#1}}}

\newcommand{\Sxn}[1]{\ensuremath{S_{x-}^{#1}}}

\newcommand{\Snx}[1]{\ensuremath{S_{+x}^{#1}}}

\newcommand{\Sxs}[1]{\ensuremath{S_{x*}^{#1}}}

\newcommand{\bxn}[1]{\ensuremath{|B_{x-}^{#1}|}}
\newcommand{\bnx}[1]{\ensuremath{|B_{+x}^{#1}|}}

\newcommand{\bxs}[1]{\ensuremath{|B_{x*}^{#1}|}}

\newcommand{\dxn}[1]{\ensuremath{|D_{x-}^{#1}|}}
\newcommand{\dnx}[1]{\ensuremath{|D_{+x}^{#1}|}}

\newcommand{\dny}[1]{\ensuremath{|D_{+y}^{#1}|}}

\newcommand{\sxn}[1]{\ensuremath{|S_{x-}^{#1}|}}
\newcommand{\snx}[1]{\ensuremath{|S_{+x}^{#1}|}}
\newcommand{\sxs}[1]{\ensuremath{|S_{x*}^{#1}|}}

\def\RatioForMC(#1){\ensuremath{2^G\cdot #1 M}}
\def\RatioForMCh(#1){\ensuremath{2^G\cdot #1 M G}}

\newcommand*{\FULLVERSION}{}%

\usepackage{amssymb,amsmath,amsthm}
\usepackage{framed}

\newtheorem{theorem}{Theorem}

\newtheorem*{lemma*}{Lemma}

\theoremstyle{definition}

\newtheorem*{remark*}{Remark}

\usepackage{subcaption}

\makeatletter
\newcommand{\text@hyphens}{\mathcode`\-=`\-\relax}
\newcommand{\id}[1]{\ensuremath{\mathit{\text@hyphens#1}}}
\makeatother

\usepackage{multirow}

\begin{document}
\maketitle

\begin{abstract}
Motivated by applications such as stock exchanges and spectrum auctions, there is a growing interest in mechanisms for arranging trade in two-sided markets. Existing mechanisms are either not truthful%
% for one side of the market
%\citep{hirai2017polyhedral},
%[Hirai and Sato 2017]
, or
do not guarantee an asymptotically-optimal gain-from-trade%
%\citep{Blumrosen2014Reallocation},
%[Blumrosen and Dobzinski 2014]
, or
rely on a prior on the traders' valuations%
%\citep{colini2017approximately},
%[Colini-Baldeschi et al. 2017]
, or
operate in limited settings such as a single kind of good%
%\citep{SegalHalevi2018MUDA}
%[Segal-Halevi et al. 2018]
.
We extend the random market-halving technique used in earlier works to markets with multiple kinds of goods, where traders have gross-substitute valuations.
We present MIDA: a Multi Item-kind Double-Auction mechanism. It is prior-free, truthful, strongly-budget-balanced, and guarantees near-optimal gain from trade when market sizes of all goods grow to $\infty$ at a similar rate. 

%\emph{Full version: {https://arxiv.org/abs/1604.06210}} 
\end{abstract}

\section{Introduction}
\label{sec:intro}
The recent auction for radio spectrum reallocation \citep{leyton2017economics}
has incited a surge of interest in \emph{double auction} mechanisms --- mechanisms for arranging trade in two-sided markets.
Such markets differ from one-sided markets in that the valuations of both buyers and sellers are their private information, and both sides might act strategically.

An important requirement from a double-auction is \emph{efficiency}, which is measured by its \emph{gain-from-trade} (GFT) --- the total value gained by the buyers minus the total value contributed by the sellers.
As an example, 
consider a stock market with two sellers: Alice  holds a unit of stock $x$ which she values as $41$ and Bob holds a unit of stock $y$ which he values as $46$. There is a single buyer, George, who wants a unit of a single stock which can be either $x$ or $y$; he values $x$ as $47$ and $y$ as $48$. Then, if George buys $x$ from Alice the GFT is $47-41=6$ while if he buys $y$ from Bob the GFT is $48-46=2$. Therefore, an optimal double-auction mechanism would make the former trade.
If the mechanism makes the wrong trade, it attains only $1/3$ of the optimal GFT.
\footnote{
Note that the \emph{social welfare} --- the sum of agents' valuations --- is much easier to approximate than the GFT. E.g, in the above example, the wrong deal attains $\frac{48+41}{47+46}>95\%$ of the optimum.
}

A common double-auction mechanism is the \emph{Walrasian mechanism} \citep{rustichini1994convergence,babaioff2014efficiency}. It computes an \emph{equilibrium price-vector} --- a price for each good, at which the supply of each good equals its demand: the total number of units that sellers want to sell at this price equals the total number of units that buyers want to buy at this price. This mechanism attains the maximum GFT \citep[Th. 11.13]{nisan2007introduction}. Unfortunately, it is not \emph{truthful} --- traders may gain by reporting false valuations.

In fact, in a two-sided market, \emph{any} truthful mechanism that attains the maximum GFT cannot be \emph{budget-balanced (BB)} --- the market-maker has to subsidize the trade \citep{myerson1983efficient}. Therefore, it is interesting to develop double-auction mechanisms that are truthful, BB and attain an \emph{approximately} maximal GFT.
We define the \emph{competitive ratio} of a mechanism as the minimum ratio (over all utility profiles) of its GFT divided by the optimal GFT. 
The first truthful and BB approximation mechanism was presented in the seminal work of \citet{mcafee1992dominant}: Its competitive-ratio is $1-1/k$, where $k$ is the number of units traded in the optimal situation.  (i.e, its GFT is always at least $1-1/k$ of the maximum GFT). Thus, McAfee's mechanism is \emph{asymptotically optimal} --- when the market-size $k$ grows to infinity, the GFT approaches the maximum.
Another advantage of McAfee's mechanism is that it is \emph{prior-free} (PF) --- it does not require any probabilistic knowledge on traders' valuations; it works even for adversarial valuations.
Its main drawback is that it works only 
in a \emph{single-good} market when each trader can trade a \emph{single unit}.

Currently, as far as we know, no truthful mechanism attains asymptotic optimality in a two-sided market with multiple kinds of goods (see related work in \S\ref{sec:related}). 
Recently, \citet{SegalHalevi2018MUDA} presented an truthful, BB, PF and asymptotically-optimal mechanism for a \emph{single-good} market that allows \emph{many units} per trader. It is called \emph{MUDA} and is based on a random-halving scheme:
\begin{framed}
\noindent
Split the market into two sub-markets, left and right, by sending each trader to each side with probability 1/2, independently of the others.
Then, in each sub-market:
\begin{enumerate}
\item Calculate a Walrasian equilibrium price ($p^R$ at the right, $p^L$ at the left).
\item Let the traders trade at the price from the other market ($p^L$ at the right, $p^R$ at the left).
\end{enumerate}
\end{framed}
The competitive ratio of MUDA is $1-O(M\sqrt{\frac{\ln k}{k}})$, 
where $k$ is again the total number of units traded in the optimal situation and $M$ is the maximum number of units per trader. 
%Thus, its GFT approaches the optimum when $k\to\infty$.

The goal of this paper is to break the single-good barrier
and attain truthful, BB, PF and asymptotic optimality in a multi-good market. 
We show that is is possible to extend the random-halving scheme of MUDA to multi-good markets.
We call the extended version MIDA --- Multiple Item-kind Double Auction.
However, the extension is not trivial and requires several additional assumptions, presented below.

\iffalse
\begin{quote}
\emph{1. Can the random-halving scheme be applied to markets with multiple goods?
\\
2. If so, does its competitive ratio approach 1 when the market is large?
}
\end{quote}
\fi
\subsection{Truthfulness in a Multi-Good Market}
%Our answer to question 1 is a qualified ``yes'': the scheme can be applied truthfully under two assumptions.

First, since \emph{step 1} of MUDA calculates Walrasian equilibria, extending it requires that a Walrasian equilibrium exists.
\citet{gul1999walrasian} prove that a sufficient condition for the existence of a Walrasian equilibrium is that all traders' valuations satisfy a condition called \emph{gross-substitute (GS)}.

The GS condition means that, if an agent wants to buy a good $x$ and the price of another good $y$ increases, the agent still wants to buy $x$ (so the demand for good $x$ weakly increases).
A typical example of a GS market is a market for used cars: usually, people consider cars of different models to be substitutes, so when the price of one model increases, the demand for other models weakly increases.

Second, \emph{step 2} of MUDA 
lets traders trade in a price that is not an equilibrium price at their own market, so it must handle excess demand and supply.
In a single-good market this is easy since there is excess only in one side of the market, so it can be solved using randomization.
In a multi-good market this is harder since in each side we can have excess demand in some goods and excess supply in other goods.

\iffalse
As an example, suppose that in one of the sub-markets there are 100 additive car-sellers, each of whom holds one Fiat and one Subaru, and 190 unit-demand buyers, each of whom wants a single car. In the market-prices, 
each seller gains both from selling Fiat and from selling Subaru, but its gain from selling Subaru is much larger.
In the market-prices, 110 buyers prefer Fiat and 80 buyers prefer Subaru. 
If all traders report truthfully, there is excess demand in Fiat and excess supply in Subaru, so a naive extension of MUDA would let all 100 sellers sell Fiat to a random subset of the 110 Fiat-buyers, and
let all 80 Subaru-buyers buy Subaru from a random subset of the sellers.
However, this lottery is no longer truthful: the sellers now have an incentive to claim that they do not want to sell Fiat at all, since this will induce the Fiat-buyers to move to Subaru, which is more profitable for the sellers.
\fi

Currently we know to solve this problem for markets in which 
only one of the two sides can trade multiple goods.
For concreteness, we assume that the buyers can trade multiple goods while each seller sells multiple units of a single good.

Under these two assumptions, MIDA works as follows.
\begin{framed}
\noindent
Halve the market randomly (like MUDA). In each half:

1. Calculate a Walrasian equilibrium price-vector --- a price for each good-kind ($\pricea$ at the right, $\priceb$ at the left).

2. Let the traders trade at the price-vector from the other market ($\priceb$ at the right, $\pricea$ at the left) as follows:
\begin{itemize}
\item For each good-kind $x$, put all the sellers of $x$ in a queue, ordered randomly. For each $x$-seller, calculate the number of $x$-units that maximize his gain at the market price-vector; sum all these numbers to get the aggregate supply of $x$.
\item Put all buyers in a single queue, ordered randomly. For each buyer $i$ in the buyers' queue, 
calculate the bundle that maximizes his net gain at the market price-vector, among all bundles that are contained in the aggregate supply. Let the buyer buy this bundle from the first sellers in the sellers' queues.
Whenever a first seller exhausts his optimal supply, he leaves the market and the next seller in this queue becomes first.
\end{itemize}
\end{framed}

\begin{theorem}
\label{thm:strategic}
Suppose each seller sells multiple units of a single good, and all traders have  gross-substitute valuations. Then, 
MIDA is strongly budget-balanced (= has no deficit and no surplus) and truthful.
Moreover, these properties hold ex-post --- for every outcome of the randomization.
\end{theorem}
\begin{proof}
Strong budget-balance is obvious since all monetary transfers are between buyers and sellers.
We now prove that this mechanism is truthful  for every randomization outcome.

The buyers effectively play random-serial-dictatorship, which is known to be truthful. Whenever it's a buyer's turn to buy a bundle, the mechanism picks for him the best possible bundle given the available supply.

The sellers have GS valuations. 
\citet{gul1999walrasian} prove that GS valuations are \emph{submodular}.
This means that a seller gains the highest utility from selling the first unit, 
a lower utility from selling the second unit, etc. Hence, 
for each seller there is an optimal number of units to sell in the given market-prices, and it is always optimal for a seller to sell as many units as possible up to the optimal number, then leave. This is exactly what the mechanism does for them. 
\end{proof}
\begin{remark*}
(a)
MIDA is truthful only for single-good sellers.
Consider 
a seller who can sell multiple goods. 
For example, an additive seller who, in the market prices, gains 5 from selling $x$ and 3 from selling $y$ (and 8 from selling both).
MIDA has to let such a seller participate simultaneously in both seller-queues.
Suppose there is a single unit-demand buyer who gains 6 from buying $x$ and 8 from buying $y$.  If the seller reports truthfully then the buyer buys $y$. But the seller can strategize by claiming that he loses from selling $y$ (i.e, standing only in the queue of $x$-sellers); then the buyer will be forced to buy $x$ and the seller will gain more.

(b) 
MIDA is truthful only for submodular sellers.
Consider a non-submodular seller. For example, a seller who values a single unit as 10 and two units as 40. If the market price is 25 per unit, then the seller 
gains -5 from selling a single unit ans +10 from selling two units.
Suppose the first buyer in the buyers' queue wants a single unit. The best response of the seller depends on the preferences of the following buyers: if another buyer wants a unit then it is best to truthfully report a supply of 2, but if no other buyer wants a unit then it is best to untruthfully report a supply of 0.
\end{remark*}

\subsection{Asymptotic Optimality in a Multi-Good Market}
Asymptotic optimality means that, when the market is sufficiently large, the GFT approaches its maximum value.
We have to define what is meant by ``large market''. 
The definition implied by
\citet{mcafee1992dominant}
is that $k$ --- the number of units traded in the optimal situation --- approaches infinity.
This definition cannot be used as-is in a multi-good market. As an extreme example, suppose the market has two independent goods $x$ and $y$, where most GFT comes from trading a single unit of $x$, but the market-size of $y$ approaches infinity. It is clear that the competitive ratio will not approach 1 in this case. 
Therefore, we extend the ``large market'' definition as follows. 
Let $G$ be the number of different good-kinds. For any good-kind $x\range{1}{G}$, let $\kx$ be the total number of units of $x$ traded in the optimal situation. A large market is a market where for all $x$, $\kx\to\infty$.
To ensure that not all these units come from few large traders, we also assume that the number of units traded by a single person is bounded by a constant $M$ (a similar assumption is used in MUDA).

Surprisingly we found out that, even in a large market, the random halving technique might attain sub-optimal GFT. This can happen when the market-size in one good is much larger than the market-size in another good.
Appendix \ref{sec:failure} shows a concrete example 
where $G=2$ and $M=1$ but the GFT is at most $63/64$ of the maximum even if $\forall x: \kx\to\infty$.

Intuitively, 
The random halving process creates an imbalance in the number of traders in the two market-halves. Standard tail bounds tell us that, with high probability, the imbalance is in the order of square-root of the number of traders sampled. 
With one good, this imbalance is negligible as $\sqrt{k}/k \to 0$ when $k\to\infty$.
However, if we have two goods $x$ and $y$ and $\sqrt{k_x} > k_y$, then the sample deviation in $x$ might destroy all the trade in $y$. 

The example in Appendix \ref{sec:failure} is interesting, since the random-halving technique has been used successfully in various domains
\citep{Goldberg2001Competitive,Goldberg2006Competitive,devanur2015envy,balcan2008reducing,balcan2007random,devanur2009adwords}, and one could think it is an omnipotent technique. Our example shows the limitations of random-halving and illustrates the inherent difficulties in multi-good markets.

In contrast to this negative example, we prove that MIDA is asymptotically-optimal when the market sizes in all goods go to infinity in a similar rate. We define 
$\kmin := \min_x \kx$ and $\kmax := \max_x \kx$ and $c := \kmax / \kmin = $ the largest ratio between the market-sizes of different goods. Then:
\begin{theorem}
\label{thm:gain-c}
Suppose all traders have GS valuations. Then
the expected competitive ratio of MIDA is at least:
\begin{align*}
1 -
18 \cdot M^{2G+2} \cdot c \cdot  {\ln{(10 G\kmax)} \over \sqrt{\kmax} }
- o(1/\kmax)
\end{align*}
where $G$ is the number of good-kinds, $M$ the maximum number of units of a single good traded by a single person, 
$\kmax$ the largest market-size of a single good, and $c = \kmax/\kmin = $ the largest ratio between market-sizes.
\end{theorem}

Note that previously it was not known if asymptotic optimality is possible even with two goods ($G=2$). Therefore, it is encouraging that we can get asymptotic optimality for a constant number of goods.

The assumption that $c$ is bounded means that, as the population grows, the market-sizes in the different good-kinds grow at a similar rate.
It is reasonable if we believe that the ratio between market-sizes depends on some characteristics of the population that remain approximately constant when the population grows.
This is analogous to the \emph{thickness} assumption introduced e.g. by \citet{kojima2009incentives} in the context of two-sided matching.

\ifdefined\FULLVERSION
This assumption can be weakened in two ways.

First, the parameter $c$ may depend on $\kmax$. Specifically, if $c={\kmax}^{r}$ where $r<0.5$ (so $\kmin \geq {\kmax} ^ {1-r}$) then the competitive ratio guaranteed by Theorem \ref{thm:gain-c} is still $1-o(1)$.

Second, it 
\else
It
\fi
is possible to replace $c$ with a different parameter ---
$h$ --- the ratio of the largest to the smallest marginal gain from a deal in a single item. We assume that the traders' valuations are normalized such that, whenever a seller sells an item to a buyer, the marginal GFT contributed by this deal is between $1$ and $h$. This assumption
comes from the random-halving literature (e.g. \citet{Goldberg2001Competitive}). There, it is common to assume that the valuation of each buyer with positive valuation is bounded in $[1,h]$, so that each buyer contributes at most $h$ to the total welfare. Our definition is the natural generalization of this assumption to a two-sided market. 
\begin{theorem}
\label{thm:gain-h}
Suppose all traders have GS valuations. Then
the expected competitive ratio of MIDA is at least:
\begin{align*}
1 -
18 \cdot M^{2G+2} \cdot G \cdot h \cdot  {\ln{(10 G\kmax)} \over \sqrt{\kmax} }
- o(1/\kmax)
\end{align*}
where $G$ is the number of good-kinds, $M$ is the maximum number of units of a single good traded by a single person, and the marginal gain from each unit traded is in $[1,h]$.
\end{theorem}
So when $M$ and $G$ are bounded, and either $c$ or $h$ (or both) are bounded, and $\kmax\to\infty$, The GFT of MIDA approaches its maximum value.

Note that MIDA does not have to know the parameters used in Theorems \ref{thm:gain-c} and \ref{thm:gain-h} ($M, G, c, h, \kmax$, etc.);
they are used only in the analysis. The mechanism itself is prior-free.

Note also that Theorems \ref{thm:gain-c} and \ref{thm:gain-h} do not need the assumption that each seller sells a single good-kind. This assumption is needed only for the truthfulness (see Theorem \ref{thm:strategic} and the following remark).

\textbf{Paper layout.}
Related work is surveyed below. 
Formal model is presented in Section \ref{sec:model}. 
Theorems \ref{thm:gain-c} and \ref{thm:gain-h} are proved in Section \ref{sec:gain}.
Section \ref{sec:future} discusses the limitations of our results and presents directions for future work.

\subsection{Related Work} \label{sec:related}
Mechanisms for multi-unit double auction are surveyed in \citet{SegalHalevi2018MUDA}. 
Below we survey more advanced mechanisms that handle multi-good markets. 

1. \citet{feng2012tahes} present TAHES --- Truthful double-Auction for HEterogeneous Spectrum. The mechanism is inspired by the problem of re-allocating radio frequency-spectrum from primary owners to secondary users. In their setting, each seller (primary owner) has a single unit of a unique good-kind, each buyer (secondary user) has a different valuation for each kind, and each buyer wants a single unit. 
This is similar to a special case of our setting, in which all buyers have unit-demand valuations.
The authors do not analyze the competitive ratio of their mechanism. %Moreover, since an important part of their mechanism ignores the traders' valuations, it is unlikely that it can attain a high worst-case GFT.%
\ifdefined\FULLVERSION
\footnote{In fact, many double-auction papers related to frequency-spectrum reallocation do not provide a theoretic analysis of their GFT. Some of them provide simulations based on data specific to the frequency-spectrum domain, and it is not clear how they perform in other domains.}
\fi

2. \citet{Blumrosen2014Reallocation} present two  \emph{Combinatorial Reallocation} mechanisms for multi-good markets. 
The first one attains a competitive ratio of $1/[8\cdot \Theta(\log M)]$ where $M$ is the maximum number of items per single seller, but it is not prior-free since it needs to know the median value of the initial endowment of each seller. The second one is prior-free and attains a competitive ratio of 1/48. 
Both competitive ratios do not approach 1.

3. \citet{Gonen2007Generalized} present a general scheme for converting a truthful mechanism with deficit to a truthful mechanism without deficit. Their scheme can handle combinatorial double auctions, but only when traders are single-valued, e.g, when each trader has a set of desired bundles and values all these bundles the same.

4. \citet{chu2006agent} present a double-auction mechanism that can handle multiple goods. However, its truthfulness depends on substitutability conditions between buyers to buyers, sellers to sellers, and buyers to sellers. It is not clear whether these conditions hold in our setting. Moreover, they do not analyze the competitive ratio in a multi-good market.

5. \citet{colini2017approximately} present a
combinatorial double-auction mechanism in a Bayesian setting (not prior-free). In contrast to our work, they approximate the social welfare rather than the gain-from-trade. 
Note that \emph{any} mechanism that attains a fraction $\alpha$ of the optimal GFT also attains a fraction of at least $\alpha$ of the optimal social welfare \citep{brustle2017approximating,colini2017fixed}, but the opposite is not true. Moreover, 
their competitive ratios (1/4, 1/6 or 1/16) do not approach $1$. 

6. \citet{gonen2017dycom} present a combinatorial dynamic double-auction mechanism that relies on partial prior information --- the maximum and minimum values of the traders. They, too, approximate the social welfare rather than the GFT: they show by simulations that their mechanism attains about 0.5 of the maximum social welfare.

7. \citet{hirai2017polyhedral} present a double-auction mechanism for budgeted buyers, that is Pareto-optimal and truthful for the buyers, but not truthful for the sellers.

\section{Model}
\label{sec:model}
\paragraph{Traders and Valuations.}
\label{sub:valuations}
We consider a market in which some traders, the ``sellers'', are endowed with discrete goods, and other traders, the ``buyers'', are endowed with unlimited money.
There are $G$ different kinds of goods
and each trader can hold at most $M$ units of each good, for some constants $G$ and $M$.
A \emph{bundle} is a vector $X \range{0}{M}^G$.
The empty bundle is denoted by ${\emptyset} = (0,\ldots,0)$.
A \emph{price-vector} is a vector $\price = (p_1,\ldots,p_G)$. The price of a bundle $X$ is $\price \cdot X$.

Each trader $i$ has a value function $v_i$ on bundles, normalized such that $v_i(\emptyset)=0$. All traders' utilities are quasi-linear in money, so the net gain of a buyer $i$ from buying a bundle $X$ at price $\price{}$ is $u_i(X,\price) = v_i(X)-\price\cdot X$.

For each seller $j$ with initial endowment $E_j$, we define 
a \emph{sale-value function} $v_j'$ on negative bundles:
$v_j'(-X) := v_j(E_j)-v_j(E_j- X)$.
Hence the net gain of seller $j$ from selling a bundle $X$ at price-vector $\price{}$ is $u_j(-X,\price{}) = \price{}\cdot X - v'_j(-X)$.
Note that the gain from no trade is 0 both for a buyer and a seller: $\forall i,\price{}: u_i(\emptyset,\price{})=0$.

We present the gain of a trader as a sum of
\emph{marginal gains}, in which there is a term for each unit traded, ordered from good $1$ to good $G$. For example, for a buyer who buys two units of good 1 and one unit of good 2, the net gain is:
\begin{align*}
u_i([2,1],\price{}) =&
u_i([1,0],\price{}) +
\\
&
[u_i([2,0],\price{}) - u_i([1,0],\price{})] +
\\
&
[u_i([2,1],\price{}) - u_i([2,0],\price{})].
\end{align*}
\begin{sellers}
Similarly, 
 for a seller who sells two units of good 1 and one unit of good 2, the net gain is:
\begin{align*}
u_i([-2,-1],\price{}) =&
u_i([-1,0],\price{}) +
\\
&
[u_i([-2,0],\price{}) - u_i([-1,0],\price{})] +
\\
&
[u_i([-2,-1],\price{}) - u_i([-2,0],\price{})].
\end{align*}
\end{sellers}
We present each trader as a set of at most $M G$ \emph{virtual traders}, each of whom trades a single unit and enjoys the marginal utility of that unit (a similar idea was used by \citet{chawla2010multiparameter}). So the above buyer represents three virtual buyers: the first buys good 1 and gains $u_i([1,0],\price{})$, the second buys good 1 and gains $u_i([2,0],\price{}) - u_i([1,0],\price{})$, and the third buys good 2 and gains $u_i([2,1],\price{}) - u_i([2,0],\price{})$.

\paragraph{Demand and Supply.}
We assume that traders' valuations are \emph{generic}, i.e, values of different bundles and of different traders are different and linearly-independent over the integers (no linear combination with integer coefficients equals zero). 
This assumption guarantees that, given a price-vector $\pricea{}$ determined in the right market, every trader $i$ in the left market has a unique bundle that maximizes $u_i(\cdot ,\price{})$, and vice versa.%
\footnote{The genericity assumption can be dropped by using centralized tie-breaking, i.e, whenever a trader has two or more demands, the market-manager may select one of them in a way that maximizes GFT. See \citet{Hsu2016Do} for other ways to handle ties in markets.}
We call this maximizing bundle
the \emph{demand} of $i$ at $\price{}$.
If $i$ is a seller then 
his demand 
is a weakly-negative vector; we call
the absolute value of this vector the \emph{supply} of $i$ at $\price{}$.

%The \emph{aggregate-demand} at price $\price{}$ is the sum of demands of all buyers. Similarly, 
%the \emph{aggregate-supply} is the sum of supplies of all sellers.
\paragraph{Gross Substitutes.}
We assume that all traders' valuations are \emph{gross-substitute} (GS);
see \citet{Kelso1982Job} and \cite{gul1999walrasian} for formal definitions. 
Intuitively, GS means that there are no complementarities between different items, so that when the price of a single unit of a single good increases, the demands for the other units and the other goods do not decrease.
For example, suppose at price-vector $\price{}$ the demand of some buyer is $[2,1]$, e.g, two units of good x and one unit of good y. Suppose that the price of the first unit of x increases while the prices of the other units remain fixed. GS means that the buyer will still want to buy at least one unit of x and one unit of y.
\begin{sellers}
Similarly, suppose the demand of some seller is $(-2,-1)$, i.e, he wants to sell two units of good 1 and one unit of good 2.
Suppose the price of a third unit of good 1 increases while the other prices remain fixed.
GS means that the seller's demand for 
other units of good 1 will be $\geq -2$
and for good 2 it will be $\geq -1$, i.e, sell \emph{at most} 
two standard units of good 1 
and one unit of good 2.
\end{sellers}

When all traders have GS valuations, a Walrasian equilibrium exists and can be found efficiently using an ascending auction \citep{gul2000english,BenZwi2013Ascending}. Intuitively, the GS property guarantees that the auctioneer can increase the prices of over-demanded items, without creating a demand-shortage in other items, until the market converges to an equilibrium.

GS valuations are always \emph{submodular}; see \citet{gul1999walrasian} for a formal definition and a proof.
Intuitively, submodularity means that, if a trader loses some deals, its marginal gain from the other deals weakly increases. 
For example, if the buyer from above arrives at the head of the buyers' queue when only one unit of good 1 is available for sale, then its gain from this unit will be $u_i([1,0],\price{})$ which is at least as large as $u_i([2,0],\price{}) - u_i([1,0],\price{})$, and its additional gain from the unit of good 2 will be $u_i([1,1],\price{}) - u_i([1,0],\price{})$ which is at least as large as $u_i([2,1],\price{}) - u_i([2,0],\price{})$.
\begin{sellers}
Similarly, suppose the seller from above manages to sell only one unit of good 1 and one unit from good 2. Then its gain from the unit of good 1 will be 
$u_i([-1,0],\price{})$ which is at least as large as $u_i([-2,0],\price{}) - u_i([-1,0],\price{})$, and its additional gain from the unit of good 2 will be $u_i([-1,-1],\price{}) - u_i([-1,0],\price{})$ which is at least as large as $u_i([-2,-1],\price{}) - u_i([-2,0],\price{})$.
\end{sellers}

%In a single-good market, the GS condition is equivalent to submodularity. Indeed, MUDA assumes that all traders are submodular.

%In a multi-good market, GS is stronger than submodularity. the stronger assumption is needed since submodularity alone does not guarantee the existence of Walrasian equilibrium \citep{gul1999walrasian}.

\paragraph{Mechanisms.}
A \emph{mechanism} is a (randomized) function that takes the traders' valuations and returns (1) a price-vector $\price{}$, (2) for each buyer(seller) $i$, the 
bundle $X_i$ to buy(sell) at $\price{}$.
A mechanism is \emph{materially-balanced} if
for each good, the number of units bought equals the number of units sold:
$\sum_{i\in\text{Buyers}}X_i = \sum_{j\in\text{Sellers}}X_j$.

A mechanism is 
\emph{ex-post truthful}, if a trader can never increase his gain by pretending to have different valuations, even when the trader knows in advance the outcomes of all randomizations done by the mechanism.

Given a trading scenario in which each trader $i$ buys/sells a bundle $X_i$, the \emph{gain-from-trade (GFT)} is the sum of gains of all traders: $GFT := \sum_{i\in\text{AllTraders}} u_i(X_i)$.
\footnote{
In some papers, the gain-for-trade is termed \emph{trader surplus} or \emph{price improvement}
\citep{chakraborty2015price}.
}

In a materially-balanced setting, the GFT does not depend on the price-vector, since the prices are canceled in the summation; money is only transferred between buyers and sellers.

\section{Competitive-Ratio Analysis}
\label{sec:gain}
For truthfulness we had to assume that, in one side of the market, each  trader specializes in a single good.
In the present section we do not need this assumption --- both buyers and sellers can trade multiple goods and are entirely symmetric. 
We first analyze the optimal trade, then the right sub-market and finally the left sub-market.

\subsection{Optimal Trade}
By the gross-substitute assumption, there exists a Walrasian equilibrium in the global population, and it attains the maximum GFT (see Section \ref{sec:model}). 
By the genericity assumption, the equilibrium allocation is unique.  We call it the ``optimal trade'' and denote any equilibrium price-vector by $\priceo$.

For any good $x\range{1}{G}$, denote by $\Bxs{}$ the set of virtual buyers and by $\Sxs{}$ the set of virtual sellers participating in the optimal trade. By
definition of equilibrium, the numbers of virtual-traders in both groups are the same; this is the number we denoted by $\kx$:
\begin{align}
\label{eq:clearing-opt}
\forall x\range{1}{G}:
&&
\bxs{} = \sxs{} = \kx
\end{align}
We call these virtual traders the \emph{efficient traders}. We make the pessimistic assumption that all GFT in the sub-markets comes only from these efficient traders. Therefore, the GFT of our mechanism depends on the numbers of efficient traders that trade in each sub-market.

The reduction in GFT has two reasons: one is the \emph{sampling error} --- efficient buyers and sellers land in different sub-markets, so they do not meet and cannot trade. This error is easy to bound using standard tail inequalities. The second reason is the \emph{pricing error} --- the prices at the sub-market might be too high or too low, which might create imbalance in the demand and supply. Analyzing this error requires careful analysis of the equilibrium in the optimal situation vs. the equilibrium in each sub-market.

\subsection{Right Sub-Market}
In the right, we calculate an equilibrium price-vector $\pricea$. 
For each good $x\range{1}{G}$, define the sets of virtual-traders:
\begin{itemize}
\item $\Bxn{}$ is the set of efficient virtual buyers of $x$ (members of \Bxs{}) who do not demand $x$ at $\pricea$.
\item $\Sxn{}$ is the set of efficient virtual-sellers (members of $\Sxs{}$) who do not supply $x$ at $\pricea$.
\item $\Bnx{}$ is the set of inefficient virtual-buyers (not members of \Bxs{}) who demand $x$ at $\pricea$.
\item $\Snx{}$ is the set of inefficient virtual-sellers (not members of \Sxs{}) who supply $x$ at $\pricea$.
\end{itemize}
We denote $\Dnx{} := \Bnx{} \cup \Sxn{}$ and $\Dxn{} := \Bxn{} \cup \Snx{}$.
These sets $\Dnx{},\Dxn{}$ represent the pricing error ---
they are responsible for the loss of welfare due to efficient traders quitting or inefficient traders competing with the efficient ones. 
The goal of our analysis below is to bound the sizes of the sets \Dxn{} and \Dnx{}.

For any set $T$ of traders, denote by $T^R$ the subset of $T$ that is sampled to the right market and by $T^L$ the subset of $T$ sampled to the left market. By definition of the equilibrium price-vector $\pricea$, for every good $x$:
\begin{align}
\label{eq:clearing-right}
\bxs{R}-\bxn{R}+\bnx{R}
=
\sxs{R}-\sxn{R}+\snx{R}
\end{align}
In order to relate (\ref{eq:clearing-opt}) and (\ref{eq:clearing-right}), we have to relate $\Bxs{R},\Sxs{R}$ to $\Bxs{},\Sxs{}$. This is be done using the following lemma.
\begin{lemma*}
For every set $T$ of virtual-traders and integer $q>0$
\begin{align}
\label{eq:sampling-det}
\text{w.p.~~} 1-2 \e{\frac{-2 q^2}{M^2 |T|}}:
&&
\abs{|T^R|-{\frac{|T|}{2}}} < q
\end{align}
(``w.p. $x$'' is a shorthand to ``with probability at least $x$'').
\end{lemma*}
The lemma is proved using Hoeffding's inequality. The proof is standard and we omit it.

We apply this lemma $2 G$ times, to $\Bxs{}$ and $\Sxs{}$ for each $x\range{1}{G}$, and combine the outcomes using the union bound. This gives, $\forall q>0$:
\begin{align}
\label{eq:sampling-*}
&\text{w.p.}~~~ 1-4 G \e{{-2 q^2\over M^2 \kmax}}:
\\
\notag
&~~~\forall x:~~~
\abs{|\Bxs{R}|-{\kx / 2}} < q
\text{~and~}
\abs{|\Sxs{R}|-{\kx/ 2}} < q
\end{align}
Combining equations \ref{eq:clearing-opt}, \ref{eq:clearing-right} and \ref{eq:sampling-*} gives, $\forall q>0$:
\begin{align}
\label{eq:sampling-R}
&\text{w.p.}& 1-4 G \e{{-2 q^2\over M^2 \kmax}}:
\\
\notag
&~~~\forall x:&
\abs{
|\Bxn{R}\cup \Snx{R}|
-
|\Bnx{R} \cup \Sxn{R}|
} < 2 q
\\
\notag
&
\implies&
\abs{\dxn{R}-\dnx{R}} < 2 q
\end{align}
Our goal now is to get a bound on $\dnx{},\dxn{}$. We proceed in several steps.

\paragraph{\textbf{Step A}: From Bound on Difference to Bound on Sets.}
Below we fix some $q>0$ and assume that 
the suffix of
inequality (\ref{eq:sampling-R}) holds. This gives us an upper bound on $\abs{\dxn{R}-\dnx{R}}$. We need upper bounds on $\dxn{R}$ and $\dnx{R}$. 
To get it, we use a property of GS valuations called \emph{Downward Demand Flow (DDF)} \citep{SegalHalevi2016Demandflow}.
Consider two price-vectors \priceo{} and \pricea{}, and let $\pricedelta :=\pricea{}-\priceo{}=$ the vector of price-increases.
Intuitively, DDF means that a trader's demand goes down the price-change ladder --- a trader switches from demanding good $x$ at price \priceo{} to demanding good $y$ at price \pricea{}, only if $\Delta_x>\Delta_y$.

\begin{sellers}
Similarly, the supply of a seller goes up the price-change ladder --- a seller switches from offering 
good $x$ at price \priceo{} to offering good $y$ at price \pricea{}, only if $\Delta_y>\Delta_x$.
\end{sellers}
\begin{lemma*}
[DDF Lemma, \citet{SegalHalevi2016Demandflow}]
\label{def:io}
Every trader with gross-substitute valuations has the following property. Whenever the price-vector changes from $\priceo$ to $\pricea$, the trader's demand changes in the following way.

(1) If the trader stopped demanding some good $x$ with $\Delta_x\leq 0$, then he started demanding a good $y$ with $\Delta_y<\Delta_x$.

(2) If the trader started demanding some good $x$ with $\Delta_x\geq 0$, then he stopped demanding a good $y$ with $\Delta_y>\Delta_x$.
\end{lemma*}
\noindent
\ifdefined\FULLVERSION
%In this notation the following set-inequalities are true:
%
%(1) For every good  x that became cheaper ($\Delta_x\leq 0$):  \,\,\,\,\,\,\,\,\,\,\,\,\,\,\,\,\,\,\,\,\,\,\,\,$\Dxn{} \subseteq \bigcup_{y<x}{\Dny{}}$.
%Hence $\Dxn{R} \subseteq \bigcup_{y<x}{\Dny{R}}$.
%
%(2) For every good  x that became more expensive ($\Delta_x\geq 0$): \,\,\,\,\,\,$\Dnx{} \subseteq \bigcup_{z>x}{\Dzn{}}$.
%Hence $\Dnx{R} \subseteq \bigcup_{z>x}{\Dzn{R}}$.

See Figure \ref{fig:tradersets-2d-yx0} for an illustration of buyer-sets in a market where all buyers have unit demand.

\begin{figure}
	\caption{\label{fig:tradersets-2d-yx0}
		\emph{
			\normalsize			
			Buyer sets in a market with two goods (x and y) and unit-demand buyers. Sellers are not shown. Each buyer is denoted by a ball whose x-coordinate is the buyer's valuation to $x$ and y-coordinate its valuation to item $y$. Prices are denoted by points $(p_x,p_y)$.
		}
	}
	
	\emph{In the global market, the optimal price is $\priceo$. There are $\kx=5$ efficient x-buyers and $\ky=5$ efficient y-buyers.
%\begin{fullversion}
%	(there are also 5 efficient x-sellers and 5 efficient y-sellers; they are not shown).
%\end{fullversion}			
}
	\begin{center}
		\includegraphics[width=\columnwidth]{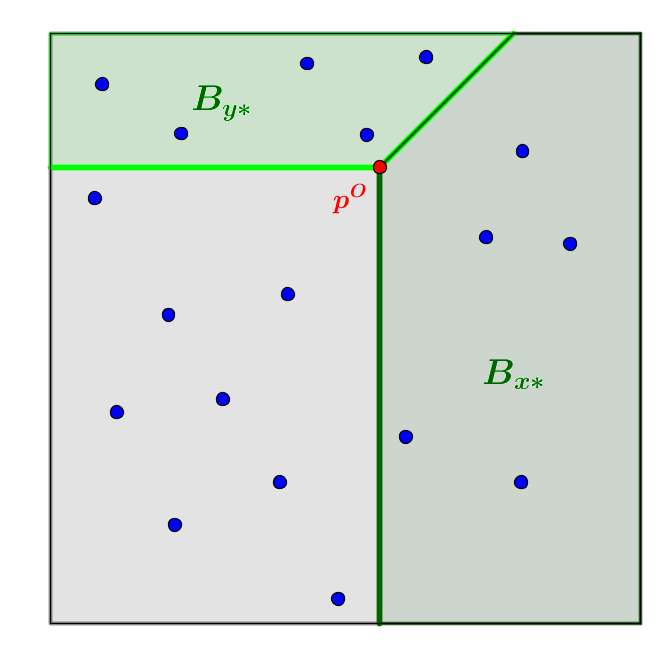}
	\end{center}
	
	\emph{The buyers are randomly divided to two markets. Buyers sampled to the right market are shown below by solid balls and the other buyers are empty balls. In the right market, the equilibrium price is $\pricea$. In this price, there are 7 y-buyers: $\Bys{R}\cup \Bny{R}$ (pentagon to the top-left of \pricea{}) and 3 x-buyers: $\Bxs{R}\setminus \Bxn{R} \cup \Bnx{R}$ (trapezoid to the right-bottom of \pricea{}). 
%	The 7 y-sellers (in $\Sys{R}\setminus \Syn{R}$) and the 3 x-sellers (in $\Sxs{R}\setminus \Sxn{R}$) are not shown.
	}
	\emph{Here, the order of price-changes is $\Delta_y<\Delta_x<0$. Item y is in the bottom of the ordering, so $\Byn{}=\emptyset$. Also, in accordance with the DDF property, $\Bxn{} \subset \Bny{}$.
	}
	\begin{center}
		\includegraphics[width=\columnwidth]{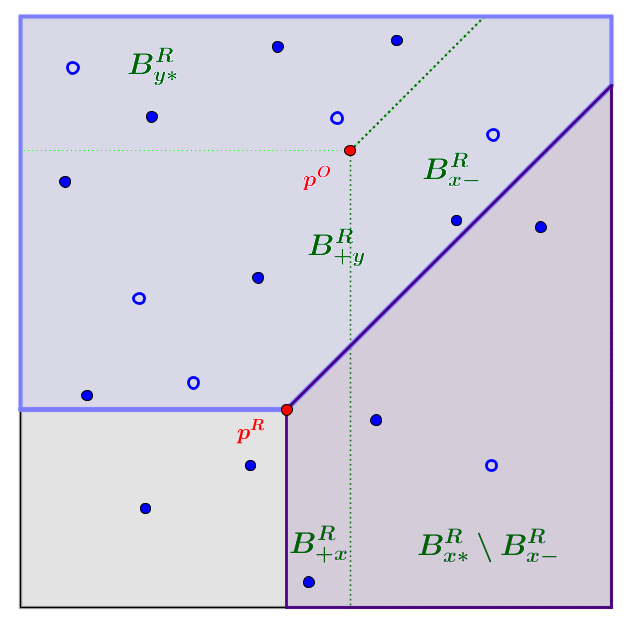}
	\end{center}
\end{figure}
%The DDF lemma implies that at least one of our $D$-sets must be empty. For example, suppose good 1 became cheaper. Then the DDF lemma implies that $D_{1-}^R=\emptyset$ since there are no goods $y$ with $\Delta_y < \Delta_1$.
%Applying (\ref{eq:sampling-R}) yields that $|D_{+1}^R|<2 q$. Applying the DDF lemma again implies that, if good 2 became cheaper then
%$D_{2-}^R\subseteq  D_{+1}^R$
%so $|D_{2-}^R|<2 q$. Applying (\ref{eq:sampling-R}) again yields that $|D_{+2}^R|<4 q$. Continuing this way by induction gives the bounds we need:
\fi

By combining the DDF lemma and \eqref{eq:sampling-R}, we can prove:
\begin{align}
\label{eq:sampling-DDF}
&\text{w.p.}~~ 1-4 G \e{{-2 q^2\over M^2 \kmax}}:
\\
\notag
&
\forall x\range{1}{G}:
~~
\dxn{R} < 2^G \cdot q
\text{~\&~}
\dnx{R} < 2^G \cdot q
\end{align}
\ifdefined\FULLVERSION
\begin{proof}
We renumber the goods in ascending order of their price change, such that $x<y$ implies $\Delta_x \leq \Delta_y$. 
In this ordering of the goods, there are zero or more goods that became cheaper, and then zero or more goods that became more expensive.
We prove \eqref{eq:sampling-DDF} for goods that became cheaper; the proof for goods that became more expensive is analogous.

We prove by induction on $x$ that $\dxn{R}$ and $\dnx{R}$ are both at most $2^{x-1}\cdot2 q$.
The base is $x=1$ --- the item with the largest price-decrease. By the DDF property,  $|D_{1-}^R|=0$. By Inequality \eqref{eq:sampling-R}, $|D_{+1}^R| < 2 q$.

For the induction step, assume the claim is true for all good-kinds $y<x$.
By the DDF property, every virtual-trader in \Dxn{R} corresponds to a virtual-trader in \Dny{R} for some $y<x$. Therefore:
\begin{align*}
	\dxn{R} &\leq \sum_{y<x}{\dny{R}}
	<
	\sum_{y<x}{2^{y-1} \cdot 2 q}
	=
	(2^{x-1}-1)\cdot 2 q
\end{align*}
By Inequality \eqref{eq:sampling-R}, $\dnx{R} < \dxn{R}+2 q = 2^{x-1}\cdot 2 q$. This concludes the induction proof.
	
Since $x-1 < G$, the lemma is proved.
\end{proof}
\else
The proof is in the full version \citep{segal2018mida}.
\fi

\paragraph{\textbf{Step B}: From a Bound on Sampled Sets to a Bound on Their Parent Sets.}
Our goal is now to derive from \eqref{eq:sampling-DDF} an upper bound on $\dxn{}$ and $\dnx{}$ --- the number of virtual traders that change their demand or supply due to the pricing error. These two sets are entirely analogous; we focus on $\Dxn{}$. 
First, we derive from (\ref{eq:sampling-det}) the following, by substituting $q\to |T|/4$:
\begin{align}
\label{eq:sampling-rev}
\forall \text{ set } T:~~
\text{w.p.~~} 1-2 \e{\frac{-|T|}{8 M^2}}:
&&
|T| < 4 |T^R|
\end{align}
We would like to apply \eqref{eq:sampling-rev} to $\Dxn{}$
and conclude that $\dxn{} < 2^G 4 q$. But we cannot do so directly 
since $\Dxn{}$ is a random-set --- it depends on the random-sampling through $\pricea$ --- while $T$ in  \eqref{eq:sampling-rev} must be a deterministic set --- independent of the random-sampling.
Our solution is to use the union bound.
Let $E_q$ be the set of all possible values of \Dxn{} with cardinality $2^G 4 q$ (so $E_q$ is a set of sets of virtual traders). 
We apply \eqref{eq:sampling-rev}
simultaneously to all sets in $E_q$ using the union bound. We get, for every integer $q$:
\begin{align*}
\text{w.p.~~} 1-2 |E_q| \e{\frac{- 2^G 4 q}{8 M^2}}:
&&
\forall T\in E_q:
~~
2^G q < |T^R|
\end{align*}
The same arguments are true for $\Dnx{}$ and for every good $x\range{1}{G}$. Combining this with \eqref{eq:sampling-DDF} gives, for every $q$:
\small
\begin{align}
\label{eq:dxndnx}
&\text{w.p.~~} 
1-4 G \e{{-2 q^2\over M^2 \kmax}}
-
4 G \e{\frac{- \ln{ |E_q| } \cdot 2^G 4 q}{8 M^2}}:
\notag
\\
&\forall x\range{1}{G}:
\dxn{} < 2^G 4q
\text{~\&~}
\dnx{} < 2^G 4q
\end{align}
\normalsize

\paragraph{\textbf{Step C}: Bounding the Number of Possible Sets.}
Our goal now is to find an upper bound on $|E_q|$. 
In fact, we will bound $|E|$ --- the number of possible values of $\Dxn{}$, regardless of their cardinality. Initially, we bound $|E|$ as a function of $n$ --- the total number of real traders that come to the auction. Then, we replace $n$ by a function of $\kmax$.

We present the set $\Dxn{}$ as a function of simpler sets. 
For every two bundles $Y$ and $Z$, denote by $R_{Y\succ Z}$ the set of traders who, at price-vector $\pricea$, prefer bundle $Y$ to bundle $Z$. 
$R_{Y\succ Z}$ is a \emph{one dimensional} random set: it contains all the traders $i$ for whom:
\begin{align*}
v_i(Y) - v_i(Z) ~~>~~ \pricea\cdot  Y - \pricea\cdot Z
\end{align*}
Therefore, for every integer $t\range{1}{n}$,
there is exactly one possible value of $R_{Y\succ Z}$ with cardinality $t$ --- it is the set of $t$ traders for whom the difference $v_i(Y) - v_i(Z)$ is the highest among all $n$ traders. Therefore the random set $R_{Y\succ Z}$ has $n$ possible values.
The total number of sets $R_{Y\succ Z}$ is at most the number of bundle-pairs, which is at most $(M^G)^2 = M^{2 G}$. Hence there are at most $n^{M^{2 G}}$ ways to choose these sets.
Once all the sets $R_{Y\succ Z}$ are determined, the set $\Dxn{}$ is completely determined too. Therefore:
\begin{align}
\label{eq:bound-E}
|E| < n^{M^{2 G}}
\end{align}
We now want to replace $n$ by an expression that depends only on $\kmax$. The reason is that $n$ --- the total number of traders in the market --- might be very large: in theory, every person on Earth might come to the auction as a ``trader'' without buying or selling anything.
Therefore, we cannot hope to get any meaningful approximation factor based on $n$.
Indeed, the approximation factors in the prior-free double-auction literature since \citet{mcafee1992dominant} are given as a function of the number of optimal deals ($k$), which is a better indicator of the actual market size.

Define a \emph{relevant trader} as a trader who trades at least one unit of at least one good in at least one price-vector --- $\priceo$ or $\pricea$.
Our arguments for bounding $|E|$ clearly hold when we use the number of relevant traders instead of $n$. 
We will now prove that, with high probability, the number of relevant traders is at most $n'$, where $n' := 10 G \kmax$. Then we will use this bound in \eqref{eq:bound-E} and get that $|E| < (n')^{M^{2 G}}$.
\begin{lemma*}
w.p.
$
1-2 \e{\frac{- 2 G^2 \kmax \ln{(8 G \kmax)}}{M^2}}
$,
the number of relevant traders is at most 
$10 G \kmax$.
\end{lemma*}
\begin{proof}
\ifdefined\FULLVERSION
\else
[Proof sketch]
\fi
By definition, the number of traders who want to trade at $\priceo$ is at most $2 G \kmax$.
It remains to count the traders who trade nothing at $\priceo$, but want to trade something at $\pricea$. 
Let $D_0$ be the set of these traders.

\ifdefined\FULLVERSION
Let $B_0$ ($S_0$) be the set of the buyers (sellers) who have zero demand (supply) at $\priceo$ and non-zero demand (supply) at $\pricea$. So $D_0 = B_0 \cup S_0$.
By the DDF property, every buyer in $B_0$ wants  only goods $x$ whose price decreased ($\Delta_x<0$), and every seller in $S_0$ offers only goods $x$ whose price increased ($\Delta_x>0$). Therefore, in the right market, each buyer in $B_0$ buys only from a seller who also sells at $\priceo$ (not in $S_0$), and every seller in $S_0$ buys only from a buyer who also buys at $\priceo$ (not in $B_0$). Hence: $|D_0^R| \leq 2 G \kmax$.

We now want to prove that, with high probability, 
$|D_0| < 8 G \kmax$. We will use the union bound and apply \eqref{eq:sampling-rev} to all possible values of $D_0$ with cardinality $8 G \kmax$.
We first have to count these possible values.
We observe that, by the DDF property, the set $B_0$ is completely determined by a union of $G$ one-dimensional sets: $\bigcup_{x=1}^G R_{\{x\} \succ \emptyset}$. This is the set of all virtual buyers who, at price-vector $\pricea$, prefer a single unit of any good over the empty set. 
Similarly, the set $S_0$ is completely determined by a union of $G$ one-dimensional sets --- the set of all virtual sellers who, at price-vector $\pricea$, prefer to sell a single unit of any good than to sell nothing.
Therefore the set $D_0$ is a union of $2 G$ one-dimensional sets. 
For every integer $t$, the number of values of $D_0$ with cardinality exactly $t$ is at most $(t+1)^{2 G - 1}$, since 
for each of the first $2 G - 1$ sets in the union, 
we have at most $t+1$ options for selecting the number of elements they will contribute to the union (once the number of elements is selected, the identity of these elements is determined, since the set is one-dimensional).

Applying \eqref{eq:sampling-rev}
simultaneously to all these sets with $t=8 G\kmax$ gives:
\scriptsize
\begin{align*}
\text{w.p.~~} 1-2(8 G\kmax+1)^{2 G - 1} \e{\frac{- 8 G \kmax}{8 M^2}}:
~~
|D_0| < 8 G \kmax
\end{align*}
\normalsize
neglecting the ``+1'' and ``-1'' and inserting the coefficient into the exponent gives the following approximation:
\small
\begin{align*}
\text{w.p.~~} 1-2 \e{\frac{- 2 G^2 \kmax \ln{(8 G \kmax)}}{M^2}}:
~~
|D_0| < 8 G \kmax
\end{align*}
\normalsize
Adding the at most $2 G\kmax$ traders who want to trade in $\priceo$ gives the claimed upper bound of $10 G \kmax = n'$.
\else
Using the DDF lemma, we can prove that their number in the right market is at most the number of traders who want to trade at $\priceo$, so $|D_0^R| \leq 2 G \kmax$. 
From this we can infer that: 
\small
\begin{align*}
\text{w.p.~~} 1-2 \e{\frac{- 2 G^2 \kmax \ln{(8 G \kmax)}}{M^2}}:
~~
|D_0| < 8 G \kmax
\end{align*}
\normalsize
so the total number of relevant traders is at most $10 G \kmax = n'$.
The proof details are deferred to the full version \citep{segal2018mida}.
\fi
\end{proof}

Substituting $n\to n'$ in \eqref{eq:bound-E} and \eqref{eq:bound-E}  in \eqref{eq:dxndnx} gives the failure probability:
\small
\begin{align*}
%\label{eq:dxndnx-failure}
4 G \e{{-2 q^2\over M^2 \kmax}}
+
4 G  \e{{M^{2 G}}\ln{n'} - \frac{2^G q}{2 M^2}}
\\
+
2 \e{\frac{- 2 G^2 \kmax \ln{(8 G \kmax)}}{M^2}}
\end{align*}
\normalsize
Denote this probability by $\mathcal{PF}(q)$. So by \eqref{eq:dxndnx}, w.p. $1-\mathcal{PF}(q)$:
\begin{align*}
\forall x\range{1}{G}:
\dxn{} < 2^G 4q
\text{~~~and~~~}
\dnx{} < 2^G 4q
\end{align*}

\subsection{Left Sub-Market}
For every good $x$, the set of efficient virtual-buyers in the left market is $\Bxs{L} = \Bxs{}\setminus \Bxs{R}$.
All these virtual-buyers want to trade at price  $\priceo$; all the virtual-buyers who are not in $\Bxn{}$ want to trade at price $\pricea$ too. Therefore, the set of efficient buyers who are sampled to the left market and want to trade at the price-vector posted for the left market is:
$\Bxs{}\setminus \Bxs{R} \setminus \Bxn{}$.
Some of these buyers might be ``wasted'' by trading with inefficient virtual-sellers; the set of inefficient virtual-sellers who want to trade $x$ at price $\pricea$ is $\Snx{}$.
All in all, the number of efficient virtual-buyers in the left submarket who want to trade at price $\pricea$ and are not wasted on inefficient sellers is at least:
\begin{align*}
\mathcal{B}_x^L := & \bxs{} - \bxs{R} - \bxn{} - \snx{}
 \\
= & \kx -  \bxs{R} - \dxn{}
\end{align*}
Similarly, the number of efficient virtual-sellers who want to trade at  $\pricea$ and are not wasted on inefficient buyers is at least:
\begin{align*}
\mathcal{S}_x^L := & \sxs{} - \sxs{R} - \sxn{} - \bnx{}
 \\
= & \kx -  \sxs{R} - \dnx{}
\end{align*}
By inequalities \eqref{eq:sampling-*} and \eqref{eq:dxndnx} from the previous subsection, 
we have for every integer $q$, with probability  $1-\mathcal{PF}(q)$:
\begin{align*}
\mathcal{B}_x^L \geq {\kx/2} - q - 2^G 4 q
\\
\mathcal{S}_x^L \geq {\kx/2} - q - 2^G 4 q
\end{align*}
Thus, in the left market, at least ${\kx/2} - q - 2^G 4 q$ efficient deals in good $x$ are done.
The same analysis is true in the right market.
Therefore, the total number of efficient deals in $x$ done in all submarkets is at least
${\kx} - 2 q - 2^G 8 q \geq \kx-2^G 9 q$.

From here we proceed to prove Theorems \ref{thm:gain-c}, \ref{thm:gain-h}.

\paragraph{Approximation Based on $c$.}
By the random ordering of the traders in each market,
the $\kx-2^G 9 q$ efficient deals executed in the sub-markets are chosen at random from the set of $\kx$ efficient deals. The same is true for every good-kind $x$.
We recall that, by submodularity, if a virtual-trader loses a deal, then the marginal-gain of other virtual-traders of the same real-trader do not decrease.  
Hence, with probability  $1-\mathcal{PF}(q)$, the expected competitive ratio is at least 
$
1 - {2^G 9 q \over \kmin}
$.
Hence, with probability 1, the expected competitive ratio is at least $1 - {2^G 9 q \over \kmin} - \mathcal{PF}(q) = $
\small
\begin{align*}
1 
&-
{2^G 9 q \over \kmin} 
- 
4 G \e{{-2 q^2\over M^2 \kmax}}
\\
&-
4 G  \e{{M^{2 G}}\ln{n'} - \frac{2^G q}{2 M^2}}
- o({1\over \kmax})
\end{align*}
\normalsize

All our analysis so far holds simultaneously for every integer $q$. Now we substitute $q ={ M^{2 G+2}\over 2^{G-1}} \cdot \ln{n} \cdot \sqrt{\kmax}$. The competitive ratio becomes:
\small
\begin{align*}
1
-&
18 M^{2G+2}  {\sqrt{\kmax}  \ln{n'} \over \kmin}
-
4 G \e{{-2 {M^{4G+2}\over 2^{G-1}} \ln^2{n'}}}
\\
-&
4 G  \e{
 {-M^{2 G}}\ln{n'} [ 
\sqrt{\kmax} - 1]} - o({1\over \kmax})
\end{align*}
\normalsize
We now substitute $n'\to 10 G \kmax$ and $\kmin \to \kmax/c$.
The latter three expressions are clearly in $o(1/\kmax)$. Therefore MIDA's competitive ratio is at least as claimed in Theorem \ref{thm:gain-c}:
\begin{align*}
1 -
18 \cdot M^{2G+2} \cdot c \cdot  {\ln{(10 G \kmax)} \over \sqrt{\kmax} }
- o(1/\kmax)
\end{align*}

\paragraph{Approximation Based on $h$.}
Recall that in each good-kind $x$, w.p. $1-\mathcal{PF}(q)$, at most ${2^G 9 q}$ deals are lost.
By assumption, each deal contributes to the GFT at most $h$. Therefore, the total loss of GFT in all goods together is at most ${2^G 9 q G h}$.
On the other hand, by assumption, each deal contributes to the GFT at least $1$. Therefore, the optimal GFT in all goods together is at least $\kmax$. Hence, w.p. $1-\mathcal{PF}(q)$, the relative loss of GFT is at most ${2^G 9 q G h \over \kmax}$. Hence, w.p. 1, the expected competitive ratio is at least $1 - {2^G 9 q G h \over \kmax}-\mathcal{PF}(q)$.
This is exactly the same expression as in the previous paragraph, except that the ${1\over \kmin}$ in the denominator is replaced by ${G h \over \kmax}$. Therefore, by the same analysis as above we get that the expected competitive ratio is at least as claimed in Theorem \ref{thm:gain-h}:
\begin{align*}
1 -
18 \cdot M^{2G+2} \cdot G \cdot h \cdot  {\ln{(10 G \kmax)} \over \sqrt{\kmax} }
- o(1/\kmax)
\end{align*}
%\fi
\section{Limitations and Future Work}
\label{sec:future}
Our goal in this work was to prove that truthful asymptotic optimality is theoretically possible even with multiple kinds of goods. 
While we proved this possibility, our work has several limitations that should be handled in future work.

(1) MIDA requires that agents have gross-substitute valuations. We do not know how to handle agents with more general valuations, e.g, submodular. The main problem is that a  Walrasian equilibrium might not exist, so we cannot use prices from one market to control trade in the other market.

(2) MIDA's truthfulness (Theorem \ref{thm:strategic}) requires that in one side of the market (e.g. the seller side), each trader specializes in a single good-kind. We do not know how to handle markets where both sides trade multiple good-kinds.
The impossibility result in 
\citet{segalhalevi2017truthful} implies that this might be a hard problem even when the prices are fixed exogenously.

(3) MIDA's asymptotic optimality
(Theorems \ref{thm:gain-c},\ref{thm:gain-h}) requires that either the parameter $c$ or the parameter $h$ (or both) are bounded.
Even with these assumptions, the competitive ratio guarantee becomes positive only for a very large market. E.g, with $M=2,G=2,c=1$, where the market-size in both goods is $k$, Theorem \ref{thm:gain-c} guarantees 
$
1 -
1152 \cdot {\ln{(20 k)} \over \sqrt{k} }
$, which becomes positive only for $k\approx 750,000,000$.

Often, the real-life performance of a mechanism is much better than its theoretical worst-case guarantee. This was shown empirically for MUDA \citep{SegalHalevi2018MUDA} and we believe this should be the case for MIDA as well. 
However, currently it is very difficult to find data on multi-good two-sided markets, since even the user-interface of such markets (e.g. in stock exchanges) usually accepts only single-good inputs.
We hope that, once multi-good double-auction mechanisms become available, two-sided markets will allow multi-good inputs, which will then enable empirical research.

%\newpage
\appendix
\section{Failure of Random Halving}
\label{sec:failure}
\newcommand{\byy}{\ensuremath{B_{yy}}}
\newcommand{\bxx}{\ensuremath{B_{xx}}}
\newcommand{\bxy}{\ensuremath{B_{xy}}}

\newcommand{\syy}{\ensuremath{S_{yy}}}
\newcommand{\sxz}{\ensuremath{S_{x\emptyset}}}

\begin{table*}
\begin{center}
\begin{tabu}{|c|c|c||c|c||c|c||c|}
\hline 
\text{Name} & 
\text{Value x} & \text{Value y} &
\text{\# Total} & 
\text{\shortstack{Optimal\\behavior}} & 
\text{\# Right} &  
\text{\# Left} & 
\shortstack{
	\text{Behavior In } $\pricea$: \\
	{\tiny $1=p^R_y\leq p^R_x\leq 6$}
}
\\
\hline
\hline \byy & 0       & 9 & $2k^4$ & \text{Buy y} & $k^4-d_1$ & $k^4+d_1$ & \text{Buy y} \\
\hline \bxy & 9       & 9 & $2k-1$ & \text{Buy x} & $k-1-d_2$ & $k+d_2$ & \text{Buy y} \\
\hline \bxx & $k^{100}$       & 0& 1      & \text{Buy x} & 0   & 1  & \text{Buy x} \\
\hline
\hline \syy & -   & 
\shortstack{
$1+i/k^5$
\\
{\small
$i=1,\ldots,2 k^4$
}
} 
& $2k^4$ &  \text{Sell y} & $k^4+k+d_3$ & $k^4-k-d_3$ &\shortstack{
$k^4+k-1-d_1-d_2$
\\
sell $y$; 
\\
$1$ is indifferent;
\\
all others do nothing.
} \\
\hline \sxz & 
\shortstack{
$6+i/k^5$   
\\
{\small
$i=1,\ldots,2 k$
}
}
& - & $2 k$   &\text{Sell x} & $k+d_4$     & $k-d_4$     & {\shortstack{
At most $1$ is indifferent;
\\
all others do nothing.
}} \\
\hline
\end{tabu}
\end{center}
\vskip -5mm
\caption{
\label{tab:failure}
Data for negative example.
}
\end{table*}
This Appendix shows that the common random-market-halving technique might fail to attain an asymptotically-optimal gain-from-trade. 
In the example, there is a two-good market
where the numbers of traders are parametrized by $k$. 
With constant probability (independent of $k$),  random-halving loses almost all the gain-from-trade even when $k\to\infty$. Hence, the competitive ratio does not approach 1 even when the market-sizes in both goods approach $\infty$.

We consider a market for medications. There are two goods (medications) denoted $y$ and $x$:
\begin{itemize}
\item Good $y$ is a medication for headache. It is 
produced by a group $\syy$ of $2 k ^4$ sellers who value it as $\approx 1$. 
More precisely, the valuation of seller $i$ in this subset is $1+i/k^5$, for $i\range{1}{|\syy|}$.
\item 
Good $x$ is a strong medication that can cure both headache and a much more lethal disease. It is produced by a group $\sxz$ of $2 k$ sellers, each of whom produces a single unit and values it as $\approx 6$:
the valuation of seller $i$ in this subset is $6+i/k^5$, for $i\range{1}{|\sxz|}$.
\end{itemize}
The groups of sellers are disjoint so all sellers are single-good and single-unit. 
There are three groups of buyers (patients).
\begin{itemize}
\item \byy{} is a large group of headache patients, each of whom wants a unit of $y$ and values it as 9.
\item \bxx{} contains a single patient who has the lethal disease and must get $x$;  he values it very highly --- $k^{100}$.
\item \bxy{} is a group of patients with a mild disease that can be cured by both $x$ and $y$; each of them wants a single unit of either $x$ or $y$ and values it as 9. 
\end{itemize}
The valuations and numbers  are summarized in Table \ref{tab:failure}. 
The first three columns show the trader sets
and their valuations.
The next two columns show the total number of traders in each group, and what they do in an optimal equilibrium. All sellers sell and all buyers buy; to balance the number of deals in $x$, all the $\bxy{}$ buyers buy $x$.
The number of efficient deals in \emph{both} good-kinds is an increasing function of $k$.

The next two columns show the number of traders from each group sampled to each sub-market. Here, the $d_j$ are random variables and the only thing we require about them is that $d_j\geq 0$. The probability of getting such a sample approaches a constant independent of $k$:
\begin{itemize}
\item The probability that in $\byy$ at least half the members fall in the left market ($d_1\geq 0$) is $1/2$; similarly for $\bxy, \sxz$.
\item The probability that the member of $\bxx$ falls in the left market is $1/2$.
\item The probability that in $\syy$ at least $k^4 + k$ members fall in the right market is less than $1/2$; however, because $k\ll \sqrt{2 k^4}$, this probability converges to $1/2$ as $k\to\infty$.
\end{itemize}
Additionally, we require that lowest-valued seller in $\sxz$ --- the seller with value $6+1/k^5$ --- falls in the right market. The probability for this is $1/2$. 
All in all, the probability of having such a sample converges to $1/64$ as $k\to\infty$.

We now calculate the equilibrium price-vector $\pricea$.
\begin{itemize}
\item The number of $y$-sellers is higher than the total number of potential $x$-buyers. Therefore, $p^R_y$ must be near $1$, so that the higher-valued sellers from $\syy$ do not sell. 
\item If $p^R_x$ is below 6 then no sellers from $\sxz$ want to sell $x$; if $p^R_x$ is at least 6 then no buyers from $\bxy$ want to buy $x$ (they prefer $y$).
In both cases, no units of $x$ are traded in the right market. Therefore $p^R_x$ is at most the value of the lowest-valued seller in the right, which is $6+1/k^5$
(in fact, $p^R_x$ can be any number between $p^R_y$ and $6+1/k^5$).
\end{itemize}
When $\pricea$ is applied in the left market, 
the price of $x$ is too low for the \sxz{} sellers so they do not sell at all. The supply of x is 0 and the \bxx{} patient cannot buy the strong medication. Almost all welfare is lost.
Since this happens with probability $\approx 1/64$, the competitive ratio is at most $\approx 63/64$ --- it does not approach 1.

This disaster \emph{cannot} happen in any of the following cases.
\begin{itemize}
\item It cannot happen when the ratio between the market-sizes of the two goods is bounded by some constant $c$. In this case, the probability that the sampling-error in $\syy$ is of the same order-of-magnitude as $\bxy$ approaches 0 as $k\to\infty$. This is consistent with our Theorem \ref{thm:gain-c}.
\item It cannot happen when the welfare contributed by a single agent is bounded by some constant $h$. In this case, the relative loss of welfare in the sampling scenario above is $h/\Omega(k)$, which approaches 0 as $k\to\infty$. This is consistent with our Theorem \ref{thm:gain-h}.
\item It cannot happen with single good-kind, even when most welfare comes from a single buyer. In this case, as long as $k$ (the number of efficient deals) is sufficiently large, there will be sufficiently many sellers willing to sell the item in a price that the important buyer is willing to pay. 
This is consistent with \citet{SegalHalevi2018MUDA}.
\end{itemize}

\section*{Acknowledgements}
This paper benefited a lot from discussions with 
Assaf Romm, Ron Adin, Simcha Haber, Ron Peretz, Tom van der Zanden, Jack D'Aurizio, Andre Nicolas, Brian Tung, Robert Israel, Clement C., C. Rose, the participants of the game-theory seminar in Bar-Ilan University, computational economics and economic theory seminars in the Hebrew university of Jerusalem, algorithms seminar in Tel-Aviv university, computer science seminar in Ariel university, and the anonymous reviewers in Theoretical Economics, SODA 2016, EC 2016, SODA 2017, EC 2017, and, of course, IJCAI 2018.

Erel was supported by ISF grant 1083/13, Doctoral Fellowships of Excellence Program and Mordecai and Monique Katz Graduate Fellowship Program at Bar-Ilan University. 
Avinatan Hassidim is supported by ISF grant 1394/16.

\fontsize{9}{9}\selectfont
%\small % 9pt
\bibliographystyle{named}
\bibliography{main}

\begin{thebibliography}{}

\bibitem[\protect\citeauthoryear{Babaioff \bgroup \em et al.\egroup
  }{2014}]{babaioff2014efficiency}
Moshe Babaioff, Brendan Lucier, Noam Nisan, and Renato Paes~Leme.
\newblock On the efficiency of the walrasian mechanism.
\newblock In {\em Proc. EC '14}, pages 783--800, 2014.

\bibitem[\protect\citeauthoryear{Balcan \bgroup \em et al.\egroup
  }{2007}]{balcan2007random}
Maria-Florina Balcan, Nikhil Devanur, Jason~D. Hartline, and Kunal Talwar.
\newblock {Random Sampling Auctions for Limited Supply}.
\newblock Technical report, Carnegie Mellon University, 2007.

\bibitem[\protect\citeauthoryear{Balcan \bgroup \em et al.\egroup
  }{2008}]{balcan2008reducing}
Maria-Florina Balcan, Avrim Blum, Jason~D. Hartline, and Yishay Mansour.
\newblock {Reducing mechanism design to algorithm design via machine learning}.
\newblock {\em J Comput. Sys. Sci.}, 74(8):1245--1270, 2008.

\bibitem[\protect\citeauthoryear{Ben-Zwi \bgroup \em et al.\egroup
  }{2013}]{BenZwi2013Ascending}
Oren Ben-Zwi, Ron Lavi, and Ilan Newman.
\newblock {Ascending auctions and Walrasian equilibrium}, July 2013.

\bibitem[\protect\citeauthoryear{Blumrosen and
  Dobzinski}{2014}]{Blumrosen2014Reallocation}
Liad Blumrosen and Shahar Dobzinski.
\newblock {Reallocation Mechanisms}.
\newblock In {\em Proc. EC}, page 617. ACM, 2014.

\bibitem[\protect\citeauthoryear{Brustle \bgroup \em et al.\egroup
  }{2017}]{brustle2017approximating}
Johannes Brustle, Yang Cai, Fa~Wu, and Mingfei Zhao.
\newblock Approximating gains from trade in two-sided markets via simple
  mechanisms.
\newblock In {\em Proc. EC '17}, 2017.

\bibitem[\protect\citeauthoryear{Chakraborty \bgroup \em et al.\egroup
  }{2015}]{chakraborty2015price}
Mithun Chakraborty, Sanmay Das, and Justin Peabody.
\newblock Price evolution in a continuous double auction prediction market with
  a scoring-rule based market maker.
\newblock In {\em AAAI}, pages 835--841, 2015.

\bibitem[\protect\citeauthoryear{Chawla \bgroup \em et al.\egroup
  }{2010}]{chawla2010multiparameter}
Shuchi Chawla, Jason~D. Hartline, David~L. Malec, and Balasubramanian Sivan.
\newblock {Multi-parameter Mechanism Design and Sequential Posted Pricing}.
\newblock In {\em Proc. STOC}, pages 311--320, 2010.

\bibitem[\protect\citeauthoryear{Chu and Shen}{2006}]{chu2006agent}
Leon~Y. Chu and Zuo-Jun~M. Shen.
\newblock {Agent Competition Double-Auction Mechanism}.
\newblock {\em Management Science}, 52(8):1215--1222, 2006.

\bibitem[\protect\citeauthoryear{Colini-Baldeschi \bgroup \em et al.\egroup
  }{2017a}]{colini2017fixed}
Riccardo Colini-Baldeschi, Paul Goldberg, Bart de~Keijzer, Stefano Leonardi,
  and Stefano Turchetta.
\newblock Fixed price approximability of the optimal gain from trade.
\newblock In {\em International Conference on Web and Internet Economics},
  pages 146--160. Springer, 2017.

\bibitem[\protect\citeauthoryear{Colini-Baldeschi \bgroup \em et al.\egroup
  }{2017b}]{colini2017approximately}
Riccardo Colini-Baldeschi, Paul~W Goldberg, Bart de~Keijzer, Stefano Leonardi,
  Tim Roughgarden, and Stefano Turchetta.
\newblock Approximately efficient two-sided combinatorial auctions.
\newblock In {\em Proc. EC '17}, pages 591--608, 2017.

\bibitem[\protect\citeauthoryear{Devanur and Hayes}{2009}]{devanur2009adwords}
Nikhil~R. Devanur and Thomas~P. Hayes.
\newblock {The Adwords Problem: Online Keyword Matching with Budgeted Bidders
  Under Random Permutations}.
\newblock In {\em Proc. EC}, pages 71--78, 2009.

\bibitem[\protect\citeauthoryear{Devanur \bgroup \em et al.\egroup
  }{2015}]{devanur2015envy}
Nikhil~R. Devanur, Jason~D. Hartline, and Qiqi Yan.
\newblock {Envy freedom and prior-free mechanism design}.
\newblock {\em JET}, 156:103--143, 2015.

\bibitem[\protect\citeauthoryear{Feng \bgroup \em et al.\egroup
  }{2012}]{feng2012tahes}
Xiaojun Feng, Yanjiao Chen, Jin Zhang, Qian Zhang, and Bo~Li.
\newblock {TAHES: Truthful double Auction for Heterogeneous Spectrums}.
\newblock In {\em Proc. INFOCOM}, pages 3076--3080. IEEE, 2012.

\bibitem[\protect\citeauthoryear{Goldberg \bgroup \em et al.\egroup
  }{2001}]{Goldberg2001Competitive}
Andrew~V. Goldberg, Jason~D. Hartline, and Andrew Wright.
\newblock {Competitive Auctions and Digital Goods}.
\newblock In {\em Proc. SODA '01}, 2001.

\bibitem[\protect\citeauthoryear{Goldberg \bgroup \em et al.\egroup
  }{2006}]{Goldberg2006Competitive}
Andrew~V. Goldberg, Jason~D. Hartline, Anna~R. Karlin, Michael Saks, and Andrew
  Wright.
\newblock {Competitive auctions}.
\newblock {\em GEB}, 55:242--269, 2006.

\bibitem[\protect\citeauthoryear{Gonen and Egri}{2017}]{gonen2017dycom}
Rica Gonen and Ozi Egri.
\newblock Dycom: A dynamic truthful budget balanced double-sided combinatorial
  market.
\newblock In {\em EXPLORE 2017 workshop, part of AAMAS'17}, pages 1556--1558,
  2017.

\bibitem[\protect\citeauthoryear{Gonen \bgroup \em et al.\egroup
  }{2007}]{Gonen2007Generalized}
Mira Gonen, Rica Gonen, and Elan Pavlov.
\newblock {Generalized Trade Reduction Mechanisms}.
\newblock In {\em Proc. EC}, pages 20--29, 2007.

\bibitem[\protect\citeauthoryear{Gul and Stacchetti}{1999}]{gul1999walrasian}
Faruk Gul and Ennio Stacchetti.
\newblock {Walrasian Equilibrium with Gross Substitutes}.
\newblock {\em JET}, 87(1):95--124, 1999.

\bibitem[\protect\citeauthoryear{Gul and Stacchetti}{2000}]{gul2000english}
Faruk Gul and Ennio Stacchetti.
\newblock {The English Auction with Differentiated Commodities}.
\newblock {\em JET}, 92(1):66--95, 2000.

\bibitem[\protect\citeauthoryear{Hirai and Sato}{2017}]{hirai2017polyhedral}
Hiroshi Hirai and Ryosuke Sato.
\newblock Polyhedral clinching auctions for two-sided markets.
\newblock {\em arXiv preprint arXiv:1708.04881}, 2017.

\bibitem[\protect\citeauthoryear{Hsu \bgroup \em et al.\egroup
  }{2016}]{Hsu2016Do}
Justin Hsu, Jamie Morgenstern, Ryan Rogers, Aaron Roth, and Rakesh Vohra.
\newblock Do prices coordinate markets?
\newblock {\em SIGecom Exch.}, 15(1):84--88, 2016.

\bibitem[\protect\citeauthoryear{Kelso and Crawford}{1982}]{Kelso1982Job}
Alexander~S. Kelso and Vincent~P. Crawford.
\newblock {Job Matching, Coalition Formation, and Gross Substitutes}.
\newblock {\em Econometrica}, 50(6):1483--1504, November 1982.

\bibitem[\protect\citeauthoryear{Kojima and
  Pathak}{2009}]{kojima2009incentives}
Fuhito Kojima and Parag~A. Pathak.
\newblock {Incentives and Stability in Large Two-Sided Matching Markets}.
\newblock {\em The American Economic Review}, 99(3):608--627, 2009.

\bibitem[\protect\citeauthoryear{Leyton-Brown \bgroup \em et al.\egroup
  }{2017}]{leyton2017economics}
Kevin Leyton-Brown, Paul Milgrom, and Ilya Segal.
\newblock Economics and computer science of a radio spectrum reallocation.
\newblock {\em Proceedings of the National Academy of Sciences},
  114(28):7202--7209, 2017.

\bibitem[\protect\citeauthoryear{McAfee}{1992}]{mcafee1992dominant}
R.~Preston McAfee.
\newblock {A dominant strategy double auction}.
\newblock {\em JET}, 56(2):434--450, 1992.

\bibitem[\protect\citeauthoryear{Myerson and
  Satterthwaite}{1983}]{myerson1983efficient}
Roger~B. Myerson and Mark~A. Satterthwaite.
\newblock {Efficient mechanisms for bilateral trading}.
\newblock {\em JET}, 29(2):265--281, 1983.

\bibitem[\protect\citeauthoryear{Nisan \bgroup \em et al.\egroup
  }{2007}]{nisan2007introduction}
Noam Nisan, Tim Roughgarden, Eva Tardos, and Vijay Vazirani, editors.
\newblock {\em {Algorithmic Game Theory}}.
\newblock Cambridge University Press, 2007.

\bibitem[\protect\citeauthoryear{Rustichini \bgroup \em et al.\egroup
  }{1994}]{rustichini1994convergence}
Aldo Rustichini, Mark~A Satterthwaite, and Steven~R Williams.
\newblock Convergence to efficiency in a simple market with incomplete
  information.
\newblock {\em Econometrica}, pages 1041--1063, 1994.

\bibitem[\protect\citeauthoryear{Segal-Halevi and
  Hassidim}{2017}]{segalhalevi2017truthful}
Erel Segal-Halevi and Avinatan Hassidim.
\newblock {Truthful Bilateral Trade is Impossible even with Fixed Prices}.
\newblock 2017.
\newblock arXiv preprint 1711.08057.

\bibitem[\protect\citeauthoryear{Segal-Halevi \bgroup \em et al.\egroup
  }{2016}]{SegalHalevi2016Demandflow}
Erel Segal-Halevi, Avinatan Hassidim, and Yonatan Aumann.
\newblock {Demand-flow of agents with gross-substitute valuations}.
\newblock {\em Operations Research Letters}, 44:757--760, 2016.

\bibitem[\protect\citeauthoryear{Segal-Halevi \bgroup \em et al.\egroup
  }{2018a}]{SegalHalevi2018MUDA}
Erel Segal-Halevi, Avinatan Hassidim, and Yonatan Aumann.
\newblock {MUDA: A Truthful Multi-Unit Double-Auction Mechanism}.
\newblock In {\em Proceedings of AAAI'18}, February 2018a.
\newblock arXiv preprint 1712.06848.

\end{thebibliography}
\end{document}